\documentclass[aip,pop,floats,reprint,superscriptaddress]{revtex4-1}
\usepackage[utf8]{inputenc}        
\usepackage{amssymb}
\usepackage{amsmath}
\usepackage[usenames]{color}
\usepackage{colortbl}
\usepackage{float} 
\usepackage{placeins} 
\usepackage{graphicx}
\usepackage{comment} 
\usepackage[pdfstartview=FitH, pdfpagemode=None, pdfpagelayout=OneColumn, colorlinks=true, linkcolor=blue, citecolor = blue]{hyperref}

 \def\bc{\begin{center}}          \def\ec{\end{center}}
 \allowdisplaybreaks
\begin{document}
 \title{Attenuation of waveguide modes in narrow metal capillaries}
 \author{P.V.Tuev}
 \author{K.V.Lotov}
 \affiliation{Budker Institute of Nuclear Physics SB RAS, 630090, Novosibirsk, Russia}
 \affiliation{Novosibirsk State University, 630090, Novosibirsk, Russia}
 \date{\today}
 \begin{abstract}

The channeling of laser pulses in waveguides filled with a rare plasma is one of promising techniques of laser wakefield acceleration. A solid-state capillary can precisely guide tightly focused pulses. Regardless of the material of the capillary, its walls behave like a plasma under the influence of a high-intensity laser pulse. Therefore, the waveguide modes in the capillaries have a universal structure, which depends only on the shape of the cross-section. Due to the large ratio of the capillary radius to the laser wavelength, the modes in circular capillaries differ from the classical $\textsl{TE}$ and $\textsl{TM}$ modes. The attenuation length for such modes is two orders of magnitude longer than that obtained from the classical formula, and the incident pulse of the proper radius can transfer up to 98\% of its initial energy to the fundamental mode. However, finding eigenmodes in capillaries of arbitrary cross-section is a complex mathematical problem that remains to be solved.
 \end{abstract}
 \maketitle

\section{Introduction}

Acceleration of particles in plasmas is now of great interest thanks to the ability of a plasma to withstand electric fields orders of magnitude stronger than in conventional radio-frequency structures. The concept develops in many directions, which differ in methods of driving the high-amplitude fields and controlling the driver.\cite{RMP81-1229,RMP85-751, NatPhot7-775,RMP90-035002,RAST9-63,RAST9-85} One of the directions is laser driven plasma wakefield acceleration in narrow capillaries. In this scheme, a short laser pulse propagates along a capillary filled with a plasma and drives a high-amplitude Langmuir wave with a phase velocity approximately equal to the light velocity $c$. The capillary prevents the diffraction of the laser pulse and extends the acceleration length either directly, by reflecting the pulse from the capillary walls,\cite{PRL82-4655,IEEE-PS28-1071,PRL92-205002,APB105-309,PoP20-083106,PRST-AB17-031303,PRST-AB17-051302,PoP19-093121,PAcc63-139,LPB19-219,PoP20-083120,PoP24-023104} or indirectly through a specific plasma profile inside.\cite{NatPhys2-696,PPCF49-B403,CRP10-130,PRL113-245002,PoP22-056703,Nat.530-190,APL99-091502,NJP9-415,NatPhys7-867,PoP16-123103,PoP16-093101,PoP25-073102}

Our study is related to laser pulse propagation in the narrowest capillaries, for which the pulse is in direct contact with the capillary walls, and the walls are either metallic or quickly ionized by the pulse. In both cases, the walls behave like a plasma. These capillaries have the potential to allow acceleration of particles to high energies using laser drivers of a modest peak power.\cite{LPB19-219} The low power is compensated by tight focusing of the driver. However, the achievable particle energy crucially depends on the pulse attenuation rate and on the structure of the waveguide modes in the capillary. Energies in the GeV range are possible only if most of the driver energy falls into a single capillary mode, and the driver pulse propagates far beyond the Rayleigh length without substantial damping on the walls.

The theory of wave propagation in metallic or ionized capillaries at conditions of interest for the wakefield acceleration has not yet been completed. The classical waveguide theory\cite{M-Schm} is not fully applicable to these conditions, as is shown in Ref.\,\onlinecite{TP49-91}. The attenuation rates are obtained either using strong simplifying assumptions\cite{TP49-91,PRE62-7168,PRE64-016404}, or numerically, in a mixture with other effects\cite{PRE64-016404,PoP13-053114}. At the same time, there is experimental evidence that the attenuation of short, high-contrast laser pulses in metallic\cite{PoP22-103111} or ionized\cite{PRE57-4899,LPB18-389} capillaries is low enough to consider the possibility of using narrow capillaries for the laser wakefield acceleration.

Any advanced study of pulse propagation needs a linear theory as a basis. In this paper, we consider the structure of capillary modes in a circular capillary, calculate the attenuation rates, and discuss the mode expansion of the incident pulse. We present exact solutions, analyze the accuracy of commonly used approximations, and show the differences between approximate and exact solutions. The depth of the study comes at the sacrifice of generality: we focus only on laser and capillary parameters of interest for wakefield acceleration. In particular, we consider copper capillaries and only those waveguide modes that can be excited by a Gaussian laser pulse. We do not discuss the mode expansion of the incident pulse under imperfect conditions, as this is considered in Refs.\,\onlinecite{JOSAB27-1400,PRE86-066411,JPP79-143,NIMA-740-273}. We also formulate the mathematical problem of finding eigenmodes in capillaries of arbitrary cross-section, which remains to be solved.

\section{Circular capillaries}\label{s2}

\subsection{Review of published results}

Consider a circular waveguide of the radius $a$ with the relative permittivity
\begin{equation}\label{m1}
    \varepsilon = \begin{cases}
     1, & r<a, \\
     \varepsilon_w, & r \geq a.
    \end{cases}
\end{equation}
Following the standard approach\cite{TP49-91,PRE65-026405,LL8}, we take the solution of Maxwell equations in the form
\begin{gather} 
E_z = e^{ikz-i\omega t+im\phi} \begin{cases}
     E_1 J_m(\varkappa_1 r), & r<a, \\
     E_2 K_m(\varkappa_2 r), & r \geq a;
    \end{cases} \\
B_z = e^{ikz-i\omega t+im\phi} \begin{cases}
     B_1 J_m(\varkappa_1 r), & r<a, \\
     B_2 K_m(\varkappa_2 r), & r \geq a;
    \end{cases} \\
E_r=\frac{(-1)^j}{\varkappa_j^2} \left(-ik\frac{\partial E_z}{\partial r} + \frac{m\omega}{c r} B_z\right), \\
E_\phi=\frac{(-1)^j}{\varkappa_j^2} \left( \frac{km}{r} E_z + \frac{i\omega}{c} \frac{\partial B_z}{\partial r}\right), \\
B_r=\frac{(-1)^j}{\varkappa_j^2} \left(-ik \frac{\partial B_z}{\partial r} - \frac{m\omega\varepsilon}{c r} E_z\right), \\
B_\phi=\frac{(-1)^j}{\varkappa_j^2} \left(\frac{km}{r} B_z - \frac{i\omega\varepsilon}{c} \frac{\partial E_z}{\partial r}\right),
\end{gather}
where $j=1,2$ corresponds to inner and outer regions, respectively,
\begin{gather}
\label{m8a}
    \varkappa_1^2=\omega^2/c^2-k^2, \\
\label{m8b}
    \varkappa_2^2=k^2-\varepsilon_w \omega^2/c^2,
\end{gather}
$J_m$ and $K_m$ are Bessel functions of the first kind and modified Bessel functions, respectively.
Continuity of $E_\phi$, $E_z$, $B_\phi$, and $B_z$ at $r=a$ yields
\begin{multline}\label{m9}
  \left(\frac{J_m^\prime}{J_m} + \frac{\varkappa_1}{\varkappa_2} \frac{K_m^\prime}{K_m} \right)
  \left( \frac{1}{\varepsilon_w}\frac{J_m^\prime}{J_m} + \frac{\varkappa_1}{\varkappa_2} \frac{K_m^\prime}{K_m} \right) = \\
  =\frac{1}{\varepsilon_w}\left( \frac{mkc}{\omega \varkappa_1 a} \right)^2 \left( 1 + \frac{\varkappa_1^2}{\varkappa_2^2} \right)^2,
\end{multline}
where primes denote derivatives with respect to arguments, and arguments of Bessel functions are: $J_m (\varkappa_1 a)$, $J_m^\prime (\varkappa_1 a)$, $K_m (\varkappa_2 a)$, and $K_m^\prime (\varkappa_2 a)$.

To solve equation (\ref{m9}) we need to specify $\varepsilon_w$. Both metals and quickly ionized solid walls are usually characterized by the Drude formula\cite{AM}
\begin{equation}\label{m10}
	\varepsilon_w (\omega) = 1 + i\frac{\omega_p^2 \tau}{\omega (1-i\omega \tau)},
\end{equation}
where $\omega_p^2 = 4 \pi n_e e^2/m = 4 \pi \sigma_0/\tau$ is the plasma frequency of conduction electrons, $n_e$ is their density, $e$ and $m$ are electron charge and mass, $\sigma_0$ is the conductivity, and $\tau$ is the electron collision frequency in the medium. The formula (\ref{m10}) correctly describes the reflection of short high-power laser pulses from various materials, which behave as a ``universal plasma mirror'' at high intensities.\cite{PRL75-252}

To be specific, we consider solutions of Eqs.(\ref{m9}) and (\ref{m10}) in the parameter area of the discussed experiments on laser wakefield acceleration.\cite{PoP22-103111,LPB19-219} In particular, we take the laser wavelength $\lambda=850$\,nm and the copper capillary of radius $a = 15\,\mu$m. The electric field of the incident laser pulse has the same direction over the entire cross-section, which corresponds to azimuthal modes with $|m|=1$ in cylindrical coordinates and the ratio
\begin{equation}\label{m12a}
    E_r = -i m E_\phi
\end{equation}
between field components. Excited capillary modes must have the same azimuthal dependence, so we pay most attention to $|m|=1$ modes.

Since $a \gg \lambda$, low-order waveguide modes are almost plane waves and similarly have \mbox{$k \approx \omega/c$}. The copper at high frequencies is characterized by $\sigma_0 \approx 1.6 \times 10^{17}\,\text{s}^{-1}$ and $\tau \approx 1.3 \times 10^{-14}$\,s.\cite{TP49-91} For these values,
\begin{equation}\label{m11}
    \varepsilon_w \approx -30 + 1.1i, \quad |\varepsilon_w| \gg 1, \quad \varkappa_2 \gg \omega/c \gg 1/a,
\end{equation}
which means that the perturbation penetrates the walls a short distance. In this case, the Leontovich boundary conditions\cite{LL8} are commonly used, which relates tangential components of the electric and magnetic fields on the walls:
\begin{equation}\label{m12}
    E_\phi = \zeta B_z, \qquad E_z = -\zeta B_\phi,
\end{equation}
where $\zeta = 1/\sqrt{\varepsilon_w}$ is the surface impedance. These conditions lead to the dispersion relation\cite{TP49-91}
\begin{equation}\label{m14}
  \left(\frac{J_m^\prime}{J_m} - \frac{i \varkappa_1 c \zeta}{\omega} \right)
  \left( \frac{J_m^\prime}{J_m} - \frac{i \varkappa_1 c}{\omega \zeta} \right)
  =\left( \frac{mkc}{\omega \varkappa_1 a} \right)^2.
\end{equation}
As can be seen from the comparison of Eqs.\,(\ref{m9}) and (\ref{m14}), using the conditions (\ref{m12}) is equivalent to the large-argument approximation for the modified Bessel functions ($K_m^\prime/K_m \approx -1$), neglecting $k^2$ in Eq.\,(\ref{m8b}), and neglecting the ratio $\varkappa_1^2/\varkappa_2^2$ in the right-hand side of Eq.\,(\ref{m9}).

With high wall conductivity, there are two small parameters in the problem: the impedance $|\zeta|$ and the ratio $\varkappa_1 c/\omega$. Depending on their ratio, the solutions of Eq.\,(\ref{m14}) take qualitatively different forms. In the case
\begin{equation}\label{m16a}
    |\zeta| \ll \varkappa_1 c/\omega
\end{equation}
(very high conductivity), the problem reduces to the classical result of the waveguide theory:\cite{LL8} there are two groups of modes, \textsl{TM} modes with
\begin{equation}\label{m15}
  J_m(\varkappa_1 a) = 0, \qquad E_z \neq 0, \ B_z \equiv 0
\end{equation}
and \textsl{TE} modes with
\begin{equation}\label{m16}
  J_m'(\varkappa_1 a) = 0, \qquad E_z \equiv 0, \ B_z \neq 0.
\end{equation}
The wave amplitude attenuates as $e^{-\alpha z}$ with
\begin{gather}\label{m18}
  \alpha = \frac{\omega \text{Re} (\zeta)}{kac}, \\
  \label{m18a}
  \alpha = \frac{c \varkappa_1^2  \text{Re} (\zeta)}{\omega ka} \left( 1 + \frac{m^2 \omega^2}{c^2 \varkappa_1^2 (a^2 \varkappa_1^2 - m^2)} \right)
\end{gather}
for \textsl{TM} and \textsl{TE} modes, respectively.

In the case
\begin{equation}\label{m20a}
    |\zeta| \gg \varkappa_1 c/\omega
\end{equation}
(short wavelength or large capillary radius), the solutions for $m \neq 0$ are circularly polarized waves\cite{TP49-91} with
\begin{equation}\label{m19}
    J_{m \pm 1}(\varkappa_1 a) = 0, \qquad \vec{B} = \pm i \vec{E}, \qquad E_r = \pm i E_\phi
\end{equation}
and the attenuation rate
\begin{equation}\label{m20}
    \alpha = \frac{\varkappa_1^2 \text{Re} (\zeta)}{2 k^2 a |\zeta|^2}.
\end{equation}
If $m>0$, then the solutions corresponding to the upper and lower signs in \eqref{m19} are called \textsl{L} and \textsl{R} modes, respectively.\cite{TP49-91} Only \textsl{R} modes comply with the requirement \eqref{m12a} and can be efficiently excited by a Gaussian pulse. The complementary \textsl{R} modes with left circular polarization also exist and correspond to $m<0$ and upper signs in \eqref{m19}.

As $|\zeta|$ increases, the $\textsl{TM}_{mn}$ modes continuously transform into $\textsl{R}_{mn}$ modes, where subscripts $m$ and $n$ denote azimuthal and radial mode numbers.\cite{TP49-91} The modes $\textsl{TE}_{mn}$ transform into $\textsl{L}_{m,n-1}$, and the mode $\textsl{TE}_{m1}$ vanishes.

\begin{figure}[b]
\includegraphics[width=197bp]{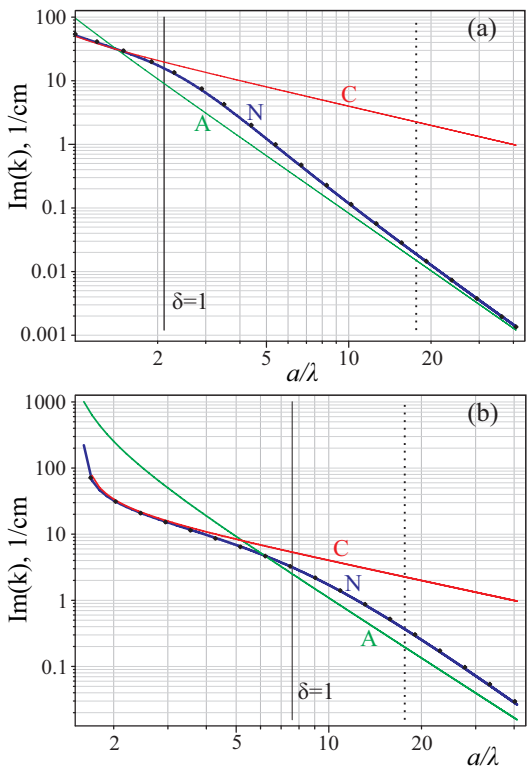}
\caption{ Attenuation rate for modes $\textsl{R}_{11}$ or $\textsl{TM}_{11}$ (a) and $\textsl{R}_{13}$ or $\textsl{TM}_{13}$ (b) calculated from classical formula (\ref{m18}) (curves `C'), approximate expression (\ref{m20}) (``A''), and numerically solved Eq.\,(\ref{m9}) (``N''). Black dots on curve ``N'' are obtained by solving Eq.\,(\ref{m14}). Thin vertical lines mark the boundary between approximations ($\delta=1$), dotted vertical lines show the considered parameter set.}\label{fig1-Imk}
\end{figure}

\subsection{Attenuation rate and mode structure}

\begin{figure}[b]
\includegraphics[width=200bp]{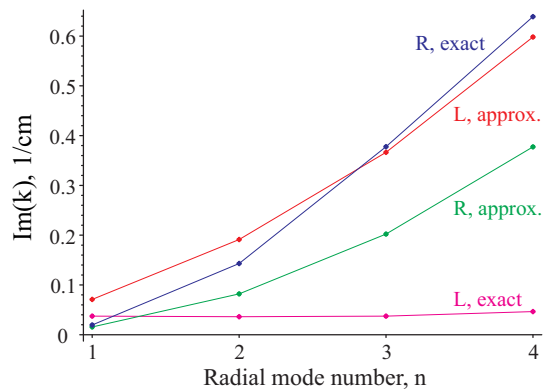}
\caption{ Attenuation rates obtained numerically (exact) and with approximation \eqref{m20} (approx.) for various modes and the baseline parameter set. }\label{fig4-difference}
\end{figure}

\begin{figure}[b]
\includegraphics[width=197bp]{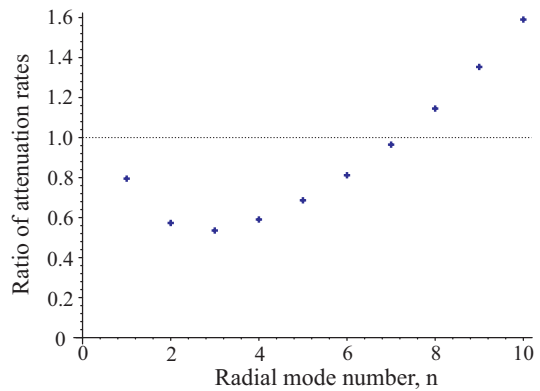}
\caption{ The ratio of attenuation rates obtained approximately and numerically for various \textsl{R} modes and the baseline parameter set. }\label{fig2-error}
\end{figure}

For the considered set of parameters, the condition \eqref{m20a} is fulfilled, but without a large margin, even for the lowest mode ($\textsl{R}_{11}$) with $\varkappa_1 a \approx 2.40483$:
\begin{equation}\label{m22}
    \zeta \approx 0.0032 - 0.18i,  \qquad  \delta \equiv \frac{\varkappa_1 c}{|\zeta|\omega} \sim 0.12.
\end{equation}
This raises the question of how accurate the approximate attenuation rate \eqref{m20} is. To answer, we compare the exact numerical solution of Eq.\,\eqref{m9} and its approximations for various ratios $a/\lambda$ (Fig.\,\ref{fig1-Imk}) and for various radial mode numbers $n$ (Fig.\,\ref{fig4-difference} and Fig.\,\ref{fig2-error}). For \textsl{R} modes with a low $n$, the approximate expression \eqref{m20} always underestimates attenuation. Although the graphs in Fig.\,\ref{fig1-Imk} are close, this is a logarithmic scale, and the difference is quite noticeable. For the lowest order radial mode ($\textsl{R}_{11}$), which should propagate for a long distance, the formula (\ref{m20}) gives an error of about 20\%. The attenuation rates for higher order \textsl{R} modes, which should quickly decay, are correct within a factor of two (Fig.\,\ref{fig2-error}), and the error almost does not decrease as the expansion parameter $\delta$ becomes smaller [Fig.\,\ref{fig1-Imk}(b)]. The numerical solution of Eq.\,\eqref{m14} almost coincides with the solution of Eq.\,\eqref{m9} (Fig.\,\ref{fig1-Imk}), so the error arises due to the approximation \eqref{m20a} rather than due to the use of Leontovich conditions \eqref{m12}. Curiously, the approximate expression is correct for \textsl{R} modes with $n\sim 7$, for which $\delta \sim 1$, and the inequality \eqref{m20a} is not valid. 


\begin{figure*}[tbh]
\includegraphics[width=447bp]{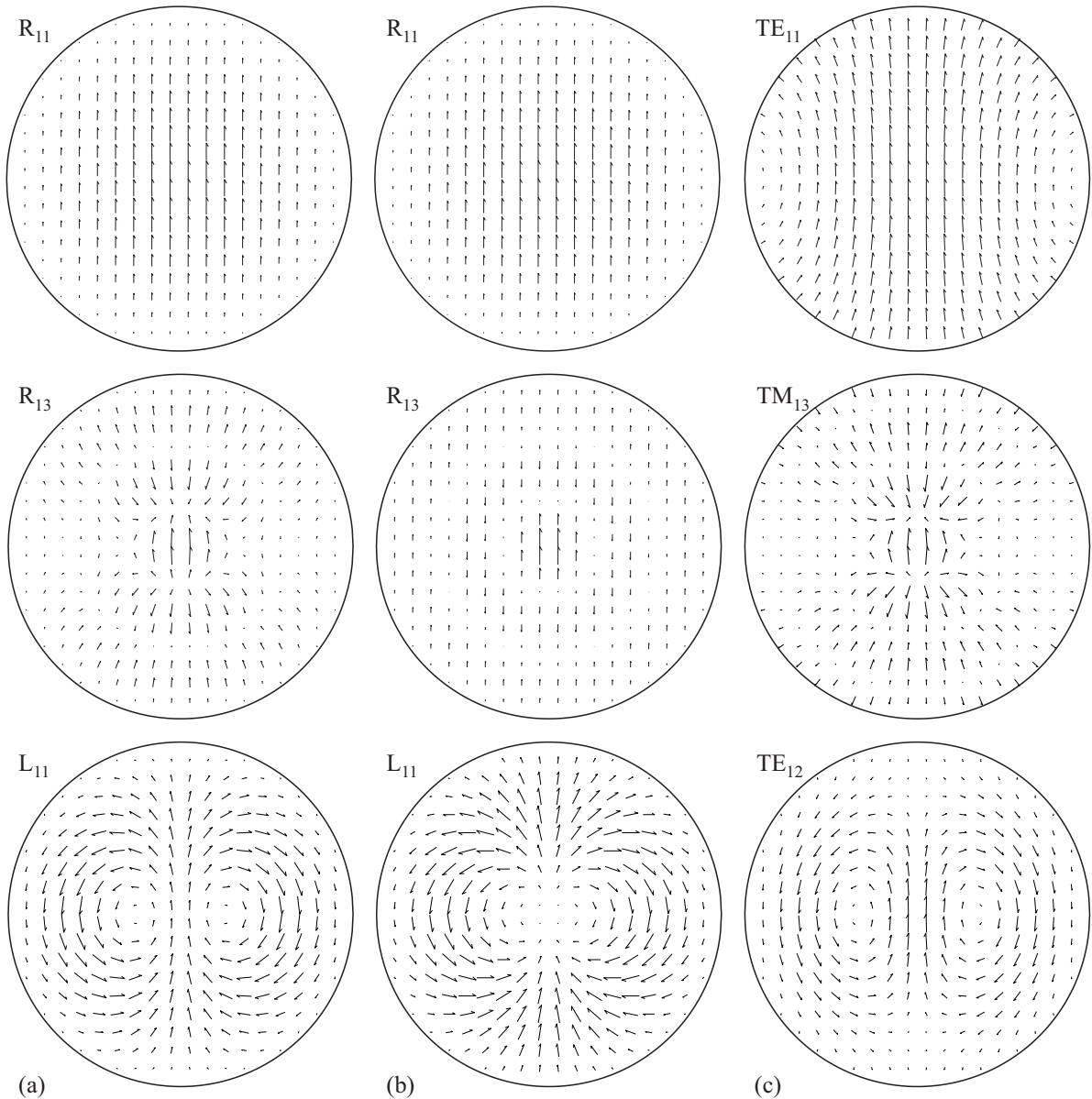}
\caption{ Transverse electric fields for different modes, calculated by the exact equations (\ref{m9}) (a) and with approximations (\ref{m20a}) (b) and (\ref{m16a}) (c).}\label{fig3-portraits}
\end{figure*}

The difference between the discussed solutions is visible in the mode structure, but only for $n>1$ (Fig.\,\ref{fig3-portraits}). The lowest order modes $\textsl{R}_{11}$ and $\textsl{TE}_{11}$, which contain most of the incident energy in the corresponding limiting cases, look similar. The main difference is that in the limit \eqref{m20a} there is no electric field on the walls. The absence (or, to be precise, a very low value) of the field on the wall explains the low attenuation rate at $\delta \ll 1$. For higher modes (the second row in Fig.\,\ref{fig3-portraits}), the exact solution contains features of both approximations: the field vectors are noncollinear, as in \textsl{TM} modes, and the field vanishes on the walls, as \eqref{m19} implies. The most significant difference is observed for \textsl{L} modes (the third row in Fig.\,\ref{fig3-portraits}): the exact solution is visually closer to the \textsl{TE} mode and, consequently, can be excited with comparable efficiency.

\begin{figure}[tb]
\includegraphics[width=200bp]{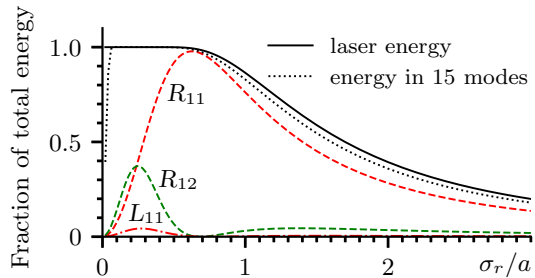}
\caption{ Laser energy falling into separate waveguide modes in relation to the radius of the incident pulse. The black solid line is the total laser energy entering the capillary. The black dotted line is the energy summed over the first fifteen modes.}\label{fig5-decomposition}
\end{figure}

To study the excitation of separate modes by incident radiation, we consider a linearly polarized laser pulse with a Gaussian field distribution at the entrance to the capillary:
\begin{align}\label{m23}
    E_y = E_0 e^{-r^2/\sigma_r^2}, \qquad E_x = 0.
\end{align}
For a wide laser pulse ($\sigma_r \gg \lambda$), we can neglect the longitudinal component of the laser electric field and calculate the energy fraction $C_\text{mode}$ that falls into a mode by integrating over the capillary cross-section $S$:
\begin{equation}\label{m24}
       C_\text{mode} = \frac{\left(\int E_y E_{\text{mode},y} \,dS \right)^2}{\int E_y^2 \, dS \int E_\text{mode}^2 \, dS},
\end{equation}
where $\vec{E}_\text{mode}$ is the transverse electric field for a $\textsl{R}_{1n}$ or $\textsl{L}_{1n}$ mode. 

At $\sigma_r / a \approx 0.64$, up to 98\% of the initial energy goes into the weakly damped $\textsl{R}_{11}$ mode (Fig.\,\ref{fig5-decomposition}). For smaller pulse radii, some energy goes into the $\textsl{L}_{11}$ mode. Since it is also weakly damped (Fig.\,\ref{fig4-difference}), its excitation can affect the wakefields inside the capillary.

\section{General case}\label{s4}

In the general case, there is no universal recipe for finding the mode structure and attenuation rates. Even the simplified approach presents serious difficulties, as we show in what follows.

The transverse structure of waveguide modes can be expressed as\cite{LL8}
\begin{gather} 
    \label{m27} \vec{E}_\perp = \frac{ik}{\varkappa^2} \nabla_\perp E_z + \frac{i \omega}{c\varkappa^2} \left[ \nabla_\perp B_z \times \vec{e}_z \right], \\
    \label{m28} \vec{B}_\perp = \frac{ik}{\varkappa^2} \nabla_\perp B_z - \frac{i \omega}{c\varkappa^2} \left[ \nabla_\perp E_z \times \vec{e}_z \right],
\end{gather}
where the subscripts $\perp$ denote transverse components of vectors, $\vec{e}_z$ is the unit vector along the capillary, and
\begin{equation}\label{m29}
    \varkappa^2=\omega^2/c^2-k^2.
\end{equation}
As we saw with circular waveguides, there is no noticeable difference between the exact solution and the solution with Leontovich boundary conditions
\begin{equation}\label{mg2}
    \vec{E}_\tau = \zeta \left[ \vec{n} \times \vec{B}_\tau \right],
\end{equation}
where $\vec{n}$ is the outer normal to the capillary wall, and the subscripts $\tau$ denote tangential components of vectors. Consequently, we can use these conditions also for other waveguide shapes. The longitudinal field components must satisfy\cite{LL8}
\begin{equation}\label{mg1}
    (\Delta_\perp + \varkappa^2) E_z = 0, \quad
    (\Delta_\perp + \varkappa^2) B_z = 0,
\end{equation}
where $\Delta_\perp$ is the two-dimensional Laplacian. Substituting expressions \eqref{m27}--\eqref{m28} for transverse field components into the Leontovich boundary conditions \eqref{mg2} yields
\begin{gather}
    \label{m34}
    \frac{\varkappa}{k \zeta} E_z = \frac{i}{\varkappa} \left( \frac{\partial B_z}{\partial \tau} + \frac{\omega}{k c} \frac{\partial E_z}{\partial n} \right), \\
    \label{m35}
    \frac{\varkappa \zeta}{k} B_z = -\frac{i}{\varkappa} \left( \frac{\partial E_z}{\partial \tau} - \frac{\omega}{k c} \frac{\partial B_z}{\partial n} \right),
\end{gather}
where the derivatives are taken along directions of $\vec{n}$ and $\vec{\tau} = -\left[ \vec{e}_z \times \vec{n} \right]$.

For low-order modes and a capillary of typical transverse size $a$, we can estimate the derivatives as
\begin{equation}\label{m36}
    \frac{\partial}{\partial \tau} \sim \frac{\partial}{\partial n} \sim \frac{1}{a} \sim |\varkappa|.
\end{equation}
In the considered parameter range ($\varkappa/k_z \ll |\zeta| \ll 1$), the fields in left-hand sides of expressions \eqref{m34} and \eqref{m35} are multiplied by small parameters. Consequently, we can neglect the left-hand sides and put
\begin{equation}\label{m37}
    \frac{\partial E_z}{\partial \tau} = \frac{\partial B_z}{\partial n}, \qquad
    \frac{\partial B_z}{\partial \tau} = - \frac{\partial E_z}{\partial n}.
\end{equation}
Solving the system \eqref{mg1} with boundary conditions \eqref{m37} should yield an approximate mode structure and the corresponding values of $\varkappa$.

The problem can be re-formulated in a more formal way. To this end, we return to real-valued fields $E_z$ and $B_z$ and introduce the complex function
\begin{equation}\label{m38}
    F = E_z + i B_z
\end{equation}
for which the equation and boundary conditions are
\begin{equation}\label{m39}
    (\Delta_\perp + \varkappa^2) F = 0, \qquad
    \frac{\partial F}{\partial n} = i \frac{\partial F}{\partial \tau}.
\end{equation}
To our knowledge, a general theory of solving the Helmholtz equation with such exotic boundary conditions has not yet been developed, and there are no algorithms (including numerical ones) that would allow finding the mode structure for capillaries of an arbitrary cross-section.

\section{Summary}\label{s5}

At parameters of interest for laser wakefield acceleration, a special regime of laser pulse propagation through the capillary is realized, in which the ratio of the laser wavelength to the capillary radius is smaller than the absolute value of the surface impedance, but not much smaller. Under these conditions, capillary eigenmodes differ from those known from the classical waveguide theory. As follows from the exact solution of Maxwell equations in a circular capillary, the available approximate solutions predict attenuation rates within a factor of two. At a certain radius of the incident laser pulse, up to 98\% of the initial energy goes into the weakly damped fundamental mode. At smaller radii, the pulse also excites a slowly damped higher mode, which is not predicted by approximate models. However, finding eigenmodes in capillaries of arbitrary cross-section is a complex mathematical problem that remains to be solved. 

The data that supports the findings of this study are available within the article.

\acknowledgements

The reported study was funded by RFBR, project number 19-31-90030.

\end{document}